\documentclass{article}
\usepackage{spconf,amsmath,graphicx}

\usepackage[hidelinks]{hyperref}
\usepackage{nameref}
\usepackage{subfigure}
\usepackage{amsfonts}
\usepackage{bbm}
\usepackage{xcolor}
\usepackage{wrapfig}
\usepackage[skip=1em]{caption} 
\usepackage{capt-of}

\newcommand\citesomething[1]{\textcolor{cyan}{\textsuperscript{[citation needed]}}}

\title{Neural Music Synthesis for Flexible Timbre Control}
\name{Jong Wook Kim$^{1,2}$, Rachel Bittner$^2$, Aparna Kumar$^2$, Juan Pablo Bello$^1$}
\address{$^1$Music and Audio Research Laboratory, New York University, $^2$Spotify}

\begin{document}
\ninept
\maketitle

\begin{abstract}
	The recent success of raw audio waveform synthesis models like WaveNet motivates a new approach for music synthesis, in which the entire process --- creating audio samples from a score and instrument information --- is modeled using generative neural networks.
	This paper describes a neural music synthesis model with flexible timbre controls, which consists of a recurrent neural network conditioned on a learned instrument embedding followed by a WaveNet vocoder.
	The learned embedding space successfully captures the diverse variations in timbres within a large dataset and enables timbre control and morphing by interpolating between instruments in the embedding space.
	The synthesis quality is evaluated both numerically and perceptually, and an interactive web demo is presented.
\end{abstract}


\section{Introduction}

Musical synthesis, most commonly, is the process of generating musical audio with given control parameters such as instrument type and note sequences over time.
The primary difference between synthesis engines is the way in which \emph{timbre} is modeled and controlled.
In general, it is difficult to design a synthesizer that both has dynamic and intuitive timbre control and is able to span a wide range of timbres; most synthesizers change timbres by having presets for different instrument classes or have a very limited space of timbre transformations available for a single instrument type.

In this paper, we present a flexible music synthesizer named Mel2Mel, which uses a learned, non-linear instrument embedding as timbre control parameters in conjunction with a learned synthesis engine based on WaveNet.
Because the model has to learn the timbre -- any information not specified in the note sequence -- to successfully reconstruct the audio, the embedding space spans over the various aspects of timbre such as spectral and temporal envelopes of notes.
This learned synthesis engine allows for flexible timbre control, and in particular, timbre morphing between instruments, as demonstrated in our interactive web demo.\footnote{\url{https://neural-music-synthesis.github.io}}


\subsection{Timbre Control in Musical Synthesis}

Methods for music synthesis are based on a variety of techniques such as FM synthesis, subtractive synthesis, physical modeling, sample-based synthesis, and granular synthesis~\cite{pejrolo2017creating}.
The method of controlling timbre and the level of flexibility depends on the parameters of the exact method used, but in general, there is a trade-off between flexible timbre control over synthetic sounds (e.g. FM or subtractive synthesis) and a limited timbre control in more ``realistic'' sounds (e.g. sample-based or granular synthesis). 
Our work is aimed at achieving the best of both worlds: flexibly controlling a variety of realistic-sounding timbres.

\pagebreak

\subsection{Timbre Morphing}

`Morphing' of a sound can be generally described as making a perceptually gradual transition between two or more sounds~\cite{caetano2010morph}.
A common approach is to use a synthesis model and define sound morphing as a numerical interpolation of the model parameters.
Sinusoidal models can directly interpolate between the energy proportions of the partials~\cite{osaka1995timbre,boccardi2001sound}.
Other models use parameters characterizing the spectral envelope~\cite{slaney1996automatic,ezzat2005morphing} or psychoacoustic features for perceptually linear transition~\cite{caetano2013musical}.
A limitation of these approaches is that morphing can only be applied among the range of timbres covered by a certain synthesis model, whose expressiveness or parameter set may be limited.
To overcome this, we employ a data-driven approach for music synthesis that is generalizable to all timbres in the dataset.

\subsection{Timbre Spaces and Embeddings}

Timbre is often modeled using a timbre space~\cite{peeters2011timbre}, in which similar timbres lie closer than dissimilar timbres.
In psychoacoustics, multidimensional scaling (MDS) is used to obtain a timbre space which preserves the timbral dissimilarities measured in perceptual experiments~\cite{grey1977multidimensional,wessel1979timbre}.
Meanwhile, in music content analysis, timbre similarity is measured using computed features such as the Mel-frequency cepstral coefficients (MFCCs) \cite{logan2000mfcc},
descriptors of the spectral envelope \cite{agostini2003musical}, or hidden-layer weights of a neural network trained to distinguish different timbres~\cite{humphrey2011nlse}.
A recent method~\cite{esling2018timbrevae} used a variational autoencoder \cite{kingma2014vae} to obtain a timbre space, and unlike the above embeddings, the method is able to generate monophonic audio for a particular timbre embedding but does not consider the temporal evolution of notes such as attacks and decays.
In our work, we generate a timbre embedding as a byproduct of polyphonic synthesis, which can utilize both spectral and temporal aspects of timbres.

\subsection{Neural Audio Synthesis using WaveNet}

WaveNet~\cite{oord2016wavenet} is a generative audio synthesis model that is able to produce realistic human speech.
WaveNet achieves this by learning an autoregressive distribution which predicts the next audio sample from the previous samples in its receptive field using a series of dilated convolutions.
Tacotron~\cite{shen2018tacotron} and Deep Voice~\cite{ping2018deepvoice} are WaveNet-based text-to-speech models which first predict a Mel spectrogram from text and use it to condition a WaveNet vocoder.

There are also a few notable applications of WaveNet on music, including NSynth~\cite{engel2017neural}, an autoencoder architecture which separately encodes monophonic pitch with learned timbral features, and the universal music translation network~\cite{mor2018universal} which uses a denoising autoencoder architecture that can extract and translate between musical styles while preserving the melody.
In \cite{anonymous2019maestro}, a WaveNet is used for music synthesis conditioned directly on note sequences, while only supporting piano sounds.
Our model is similarly built around WaveNet for its synthesis capability, while using a learned embedding space to flexibly control the timbre of polyphonic music.

\pagebreak 

\section{Method}\label{sec:method}

\begin{figure}[t]
	\centering
	\includegraphics[width=\linewidth]{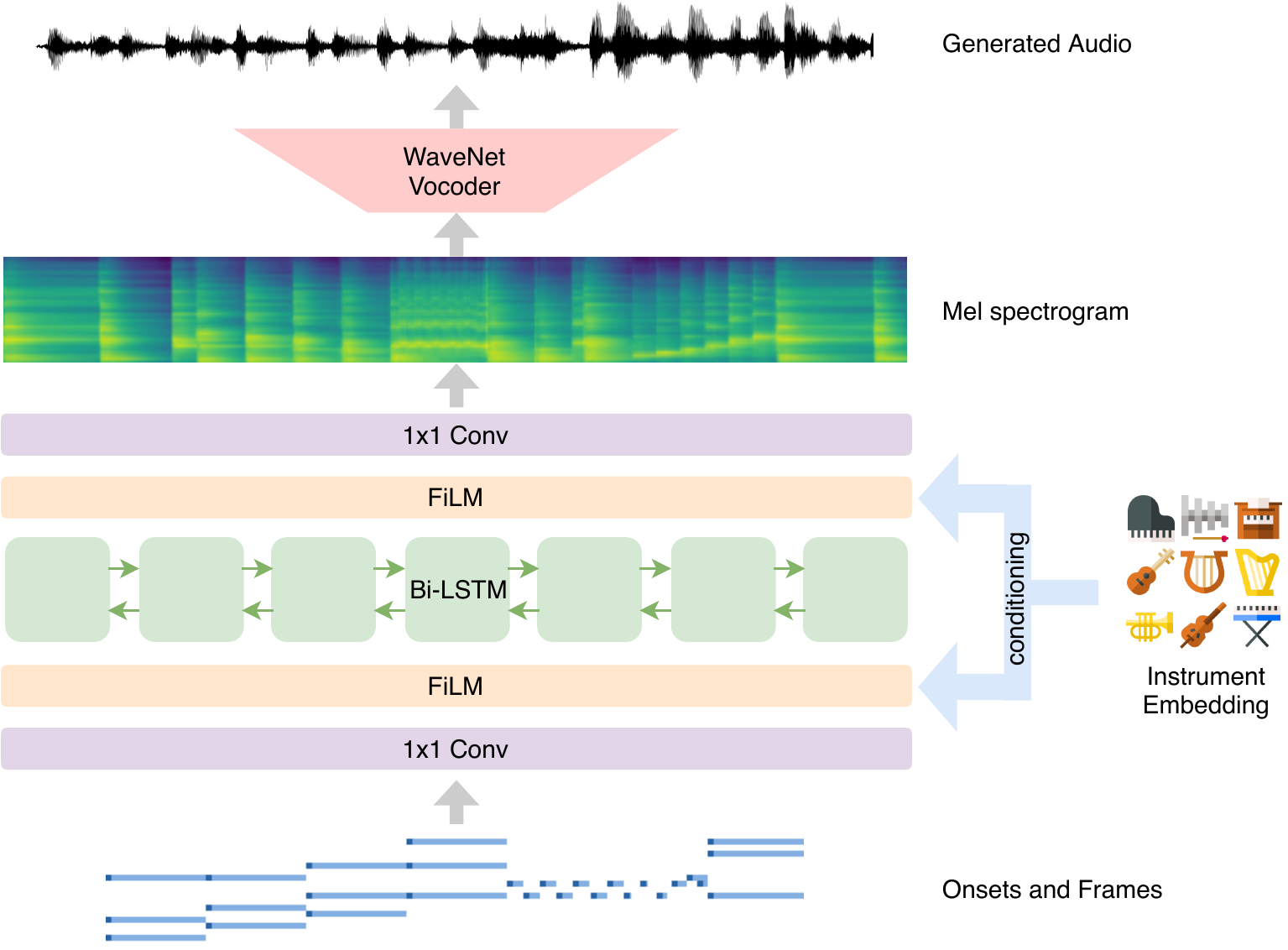}
	\caption{The overall architecture of the proposed Mel2Mel model. The note sequences and instrument embeddings are combined to predict the Mel spectrogram, which is then fed to the WaveNet vocoder.}
	\label{fig:architecture}
\end{figure}

The neural network shown in Figure \ref{fig:architecture}, dubbed Mel2Mel, concerns the task of synthesizing music corresponding to given note sequences and timbre.
The note sequences are supplied as a MIDI file and converted to a piano roll representation, which contains the note timings and the corresponding note velocities for each of the 88 piano keys.
We use a fixed step size in time, and the piano roll representation is encoded as a matrix by quantizing the note timings to the nearest time step.
The input to the neural network is a concatenation of two 88-dimensional piano roll representations, one for onsets and one for frames, comprising 176 dimensions in total:
\begin{eqnarray*}
	\mathbf{X} & =&  [~ \mathbf{X}^{\textrm{onset}} ~;~ \mathbf{X}^{\textrm{frame}} ~] \\
	\mathbf{X}^{\textrm{onset}}_{p,t} &= & v_{\textrm{~the active note}} \cdot \mathbbm{1}_{\textrm{a note at pitch $p$ is \textit{first} active at time $t$}} \\
	\mathbf{X}^{\textrm{frame}}_{p,t} &= & v_{\textrm{~the active note}} \cdot \mathbbm{1}_{\textrm{a note at pitch $p$ is active at time $t$}}
\end{eqnarray*}
\noindent where $\mathbbm{1}$ is the indicator function, and $v$ denotes the MIDI velocity scaled to $[0, 1]$.
This input representation is inspired by \cite{hawthorne2017onsets} which showed that jointly training on onsets and frames performs better than using frame information only;
similarly, we want the network to maximally utilize the onsets which have the most relevant information on the attack sounds, while still receiving the frame information.
Another reason for using both onsets and frames is that, because of the time quantization, repeated notes become indistinguishable only using $\mathbf{X}^\textrm{frame}$ when an offset is too close to the subsequent onset.

The input goes through a linear 1x1 convolution layer, which is essentially a time-distributed fully connected layer, followed by a FiLM layer, to be described in the following subsection, which takes the timbre embedding vector and transforms the features accordingly.
After a bidirectional LSTM layer and another FiLM layer for timbre conditioning, another linear 1x1 convolution layer produces the Mel spectrogram prediction.
The resulting Mel spectrogram is then fed to a WaveNet vocoder to produce the music; Mel spectrograms compactly convey sufficient information for audio synthesis and have been successfully used for conditioning WaveNet~\cite{shen2018tacotron,ping2018deepvoice}.
The use of bidirectional LSTM is justified because Mel spectrograms are constructed using a larger window than the step size, making it non-causal.
The only nonlinearities in the network are in the LSTM, and there are no time-domain convolutions except in WaveNet.

\subsection{Timbre Conditioning using FiLM Layers}

We can think of this architecture as a multi-step process that shapes a piano roll into its Mel spectrogram, by applying appropriate timbre given as side information.
A FiLM layer~\cite{perez2018film} is a suitable choice for this task, because it can represent such action of shaping using an affine transformation of intermediate-layer features.
For each instrument to model, its timbre is represented in an embedding vector $\mathbf{t}$, implemented as a learned matrix multiplication on one-hot encoded instrument labels.
A FiLM layer learns functions $f$ and $h$, which are simply linear layers mapping the timbre embedding $\mathbf{t}$ to $\gamma = f ( \mathbf{t} )$ and $\beta = h ( \mathbf{t} )$.
The affine transformation, or FiLM-ing, of intermediate-layer features $\mathbf{F}$ is then applied using feature-wise operations: $
\mathrm{FiLM}(\mathbf{F} ~|~ \gamma, \beta)
= \gamma \mathbf{F} + \beta
= f(\mathbf{t})\mathbf{F} + h(\mathbf{t}). $

At a higher level, the affine transformations learned by the FiLM layers are nonlinearly transformed by the recurrent and convolutional layers to respectively form temporal and spectral envelopes, which are two important aspects that characterize instrumental timbre.
Using the first FiLM layer is essential because the recurrent layer needs to take timbre-dependent input to apply the temporal dynamics according to the timbre, and the second FiLM layer can apply additional spectral envelope on the recurrent layer's output.

\subsection{Model Details}

The sampling rate of 16 kHz and $\mu$-law encoding with 256 quantization levels are used for all audio as in the original WaveNet paper~\cite{oord2016wavenet}.
The predicted Mel spectrograms are defined using 80 area-normalized triangular filters distributed evenly between zero and the Nyquist frequency in the Mel scale.
The STFT window length of 1,024 samples and the step size of 128 samples are used, which translate to 64 milliseconds and 8 milliseconds, respectively.
Unless specified otherwise, we use 256 channels in all hidden layers and a two-dimensional embedding space for timbre conditioning.

\medmuskip=0mu
\thinmuskip=0mu
\thickmuskip=0mu

For Mel spectrogram prediction, an Adam optimizer with the initial learning rate of 0.002 is used, and the learning rate is halved every 40,000 iterations.
The model is trained for 100,000 iterations, where each iteration takes a mini-batch of 128 sequences of length 65,536, or 4.096 seconds.
Three different loss functions are used and compared; for linear-scale Mel spectrograms $S_{\mathrm{true}}$ and $S_{\mathrm{pred}}$:
\begin{align}
\footnotesize \textbf{abs MSE} & ~=~ \mathbb{E} \left [ \left ( S_{\mathrm{true}} - S_{\mathrm{pred}} \right )^2 \right ]
\label{eqn:abs-loss} \\
\footnotesize \textbf{log-abs MSE} & ~=~ \mathbb{E} \left [ \left ( \log S_{\mathrm{true}} - \log S_{\mathrm{pred}} \right )^2 \right ]
\label{eqn:log-abs-loss} \\
\footnotesize \textbf{tanh-log-abs MSE} & ~=~ \mathbb{E} \left [ \left ( \tanh \tfrac{1}{4}\log S_{\mathrm{true}} - \tanh \tfrac{1}{4}\log S_{\mathrm{pred}}  \right )^2 \right ]
\label{eqn:tanh-log-abs-loss}
\end{align}
\noindent All logarithms above are natural, and the spectrogram magnitudes are clipped at -100 dB. Prepending $\tanh$ gives a soft-thresholding effect where the errors in the low-energy ranges are penalized less than the errors close to 0 dB.

For the WaveNet vocoder, we used nv-wavenet\footnote{\url{https://github.com/NVIDIA/nv-wavenet}}, a real-time open-source implementation of autoregressive WaveNet by NVIDIA.
This implementation limits the recurrent channel size at 64 and the skip channels at 256, because of the GPU memory capacity.
A 20-layer WaveNet model was trained with the maximum dilation of 512, and the Mel spectrogram input is upsampled using two transposed convolution layers of window sizes 16 and 32 with strides of 8 and 16, respectively.
An Adam optimizer with the initial learning rate of 0.001 is used, and the learning rate is halved every 100,000 iterations, for one million iterations in total.
Each iteration takes a mini-batch of 4 sequences of length 16,384, i.e. 1.024 seconds.

\section{Experiments}\label{sec:experiments}

\subsection{Datasets}

While it is ideal to use recorded audio of real instruments as the training dataset, the largest multi-instrument polyphonic datasets available such as MusicNet~\cite{thickstun2017musicnet} is highly skewed, contains a limited variety of solo instrument recordings, and is expected to have a certain degree of labeling errors.
So we resorted to using synthesized audio for training and collected MIDI files from \texttt{www.piano-midi.de}, which are also used in the MAPS Database~\cite{emiya2010multipitch}; these MIDI files are recorded from actual performances and contain expressive timing and velocity information.
We have selected 10 General MIDI instruments shown in Figure~\ref{fig:embedding} covering a wide variety of timbres, and 334 piano tracks are synthesized for each instrument using FluidSynth with the default SoundFont from MuseScore 3.
The 334 tracks are randomly split into 320 for training and 14 for validation.
The total size of the synthesized dataset is 3,340 tracks and 221 hours.

For later experiments, we also generate a similar dataset using 100 manually selected instrument classes using a high-quality collection of SoundFonts, which contains a wide variety of timbres.

\subsection{Ablation Study on Model Design}

In this series of experiments, we examine how slight variations in the model architecture affect the performance and show that the proposed model achieves the best performance in accurately predicting Mel spectrograms.
The first two variations use either the frame data or the onset data only as the input.
The next three omit an architectural component: one of the two FiLM layers or the backward LSTM.
The last four increase the network's capacity by adding the ReLU nonlinearity after the first convolution, using kernel sizes of 3 or 5 time steps in convolutions, or adding another LSTM layer.
\begin{center}
	\setlength\tabcolsep{0.3em}
	\begin{tabular}{c|c|c}
		Variations & Train loss ($\times 10^3$) & Validation loss ($\times 10^3$)\\
		\hline
		Proposed & 4.09 $\pm$ 0.30 & \textbf{4.75 $\pm$ 0.05} \\
		\hline
		Frame input only & 5.58 $\pm$ 0.32 & 5.92 $\pm$ 0.08 \\
		Onset input only & 5.88 $\pm$ 0.37 &  6.97 $\pm$ 0.06 \\
		\hline
		First FiLM only & 4.55 $\pm$ 0.32 & 4.99 $\pm$ 0.06 \\
		Second FiLM only & 7.65 $\pm$ 0.34 & 8.76 $\pm$ 0.08 \\
		Forward LSTM only & 5.70 $\pm$ 0.43 & 5.56 $\pm$ 0.09 \\
		\hline
		ReLU activation & 3.97 $\pm$ 0.35 & 5.04 $\pm$ 0.06 \\
		3x1 convolutions & 3.66 $\pm$ 0.28 & 5.12 $\pm$ 0.08 \\
		5x1 convolutions & 3.49 $\pm$ 0.30 & 5.06 $\pm$ 0.08 \\
		2-layer LSTM & 2.98 $\pm$ 0.20 & 4.96 $\pm$ 0.12 \\
	\end{tabular}
\end{center}
The train and validation losses as defined in Equation \ref{eqn:tanh-log-abs-loss} are shown\footnote{The means and standard deviations over the model checkpoints in 90k-100k iterations are reported, to minimize the variability due to SGD.} in the table above for each variation.
Using both onsets and frames is indeed more effective than using only one of them in the input.
The first FiLM layer plays a more crucial role than the second, because only the first can help learn a timbre-dependent recurrent layer.
As expected, removing the backward LSTM also hurt the performance.

On the other hand, any variations increasing the model capacity make the model overfit and fail to generalize to validation data.
This implies that the proposed model has the optimal architecture among the tested variations, and more specifically, having the nonlinearity only in the single recurrent layer helps the model better generalize in predicting Mel spectrograms from unseen note sequences.
A possible interpretation is that the increased capacity is being used for memorizing the note sequences in the training dataset, as opposed to learning to model the timbral features independent of specific notes.

\subsection{Synthesis Quality}

\begin{figure}[t]
	\vspace{1em}
	\hspace{-5pt}
	\includegraphics[width=0.5\linewidth]{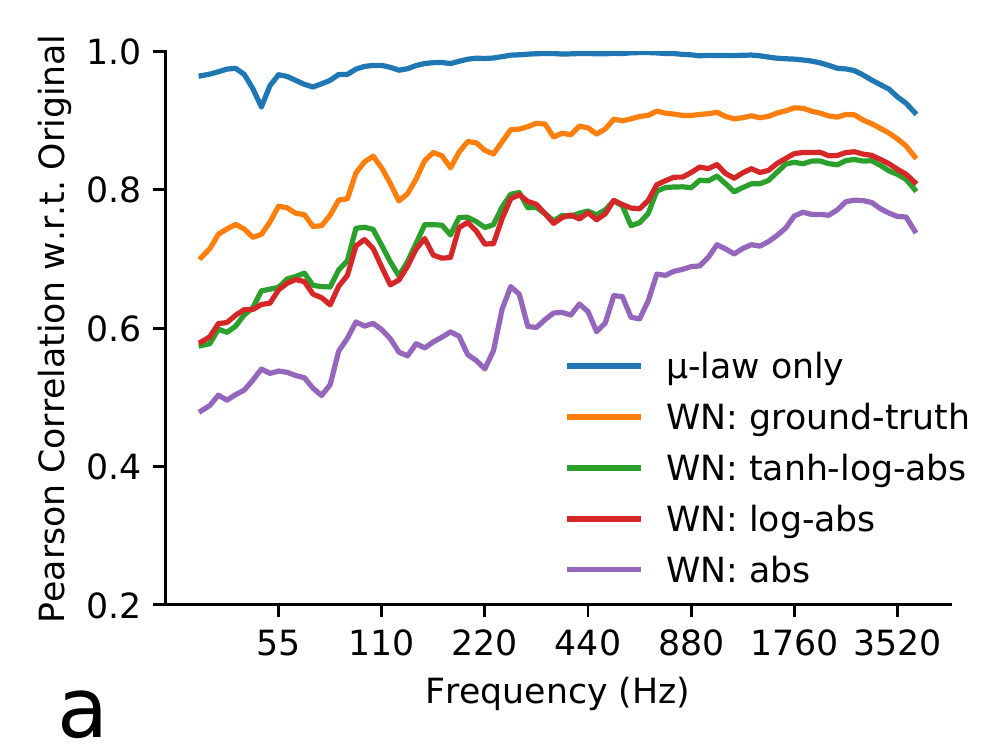}
	\includegraphics[width=0.5\linewidth]{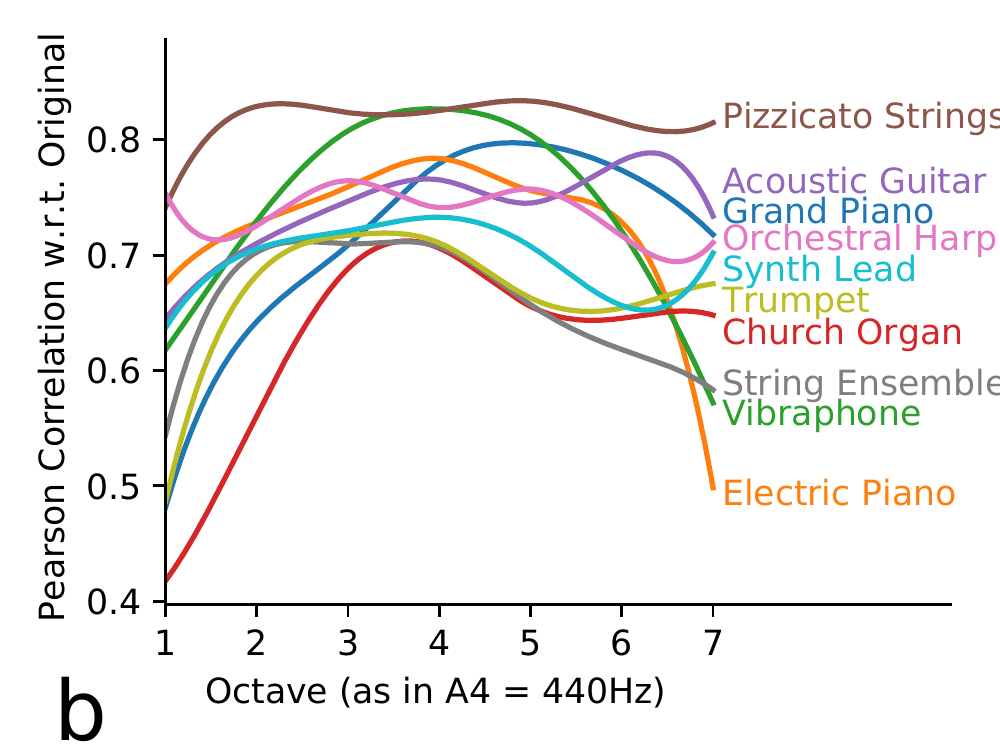}
	\caption{Pearson correlations between the reconstructed and original audio. (a) each stage of degradation. (b) per-instrument breakdown of the green curve on the left. The curves are smoothed and drawn with respect to each octave for readability.}
	\label{fig:correlations}
	\vspace{-1em}
\end{figure}

\subsubsection{Numerical Analysis of Audio Degradation}

The model goes through several stages of prediction, and each stage incurs a degradation of audio quality.
There necessarily exists some degradation caused by the $\mu$-law quantization, and WaveNet adds additional degradation due to its limited model capacity.
The generated audio is further degraded when imperfect Mel spectrogram predictions are given.
As an objective measure of audio quality degradation at each stage and for each instrument, we plot the Pearson correlations between the synthesized and original audio in Figure \ref{fig:correlations}.
To calculate and visualize the correlations with respect to evenly spaced octaves, we use 84-bin log-magnitude constant-Q transforms with 12 bins per octave starting from C1 ($\approx 32.70 \textrm{Hz}$) and 512-sample steps.
For ideal synthesis, the Pearson correlation should be close to $1$, and lower correlations indicate larger degradation from the original.

Figure \ref{fig:correlations}a shows the correlations for each stage of degradation and for different loss functions used for training the model.
The degradations are more severe in low frequencies in general, where the WaveNet model sees less number of periods of a note within its fixed receptive field length.
The orange curve showing the correlations for WaveNet synthesis using ground-truth Mel spectrograms already exhibits a significant drop from the top curve; this defines an upper bound of Mel2Mel's synthesis quality.
The lower three curves correspond to the loss functions in Equations \ref{eqn:abs-loss}-\ref{eqn:tanh-log-abs-loss}, among which the abs MSE loss clearly performs the worse than the other two which have almost identical Pearson correlation curve, indicating that the MSE loss is more effective in the log-magnitude scale.

Figure \ref{fig:correlations}b shows the breakdown of the curve corresponding to Equation \ref{eqn:tanh-log-abs-loss} into each of the 10 instruments.
There are rather drastic differences among instruments, and most instruments have low Pearson correlations in low pitches except pizzicato strings.
The reasons and implications of these trends are discussed in the following subsection, in comparison with the subjective audio quality test.

\subsubsection{Subjective Audio Quality Test}\label{sec:subjective}

We performed a crowd-sourced test asking the listeners to rate the quality of 20-second audio segments using a 5-point mean opinion score (MOS) scale with 0.5-point steps.
MOS allows a simple interface that is more suitable for non-expert listeners in a crowdsourcing setup than e.g. MUSHRA \cite{itu2011mushra} and is used as a standard approach for assessing the perceptual quality of WaveNet syntheses \cite{oord2016wavenet,oord2017parallel}.
For each of the six configurations corresponding to the curves in Figure \ref{fig:correlations}a in addition to the original audio, the first 20 seconds of the 140 validation tracks are evaluated.
The listeners are provided with a randomly selected segment at a time, and each segment is evaluated by three listeners, comprising 420 samples in each configuration.

\begin{center}
	\begin{tabular}{l|c} 
		Condition & Mean Opinion Scores \\ \hline
		Original audio & 4.301 $\pm$ 0.080 \\
		$\mu$-law encode-decoded audio & 3.876 $\pm$ 0.097 \\
		WaveNet: ground-truth Mel & 3.383 $\pm$ 0.100 \\
		\hline
		WaveNet: tanh-log-abs MSE & \textbf{3.183 $\pm$ 0.106} \\
		WaveNet: log-abs MSE & 3.019 $\pm$ 0.109 \\
		WaveNet: abs MSE & 2.751 $\pm$ 0.110 \\
	\end{tabular}
\end{center}

This table shows the mean opinion scores (MOS) and the 95\% confidence intervals, which generally follow the tendency similar to Figure \ref{fig:correlations}.
Using the soft-thresholding loss in Equation \ref{eqn:tanh-log-abs-loss} results in the best subjective quality among the three loss functions compared, more significantly so than the numerical comparison in Figure \ref{fig:correlations}.

\setlength{\intextsep}{.1em}
\setlength{\columnsep}{1em}%

\begin{wrapfigure}{r}{0.4\linewidth}
	\centering
	\includegraphics[width=\linewidth]{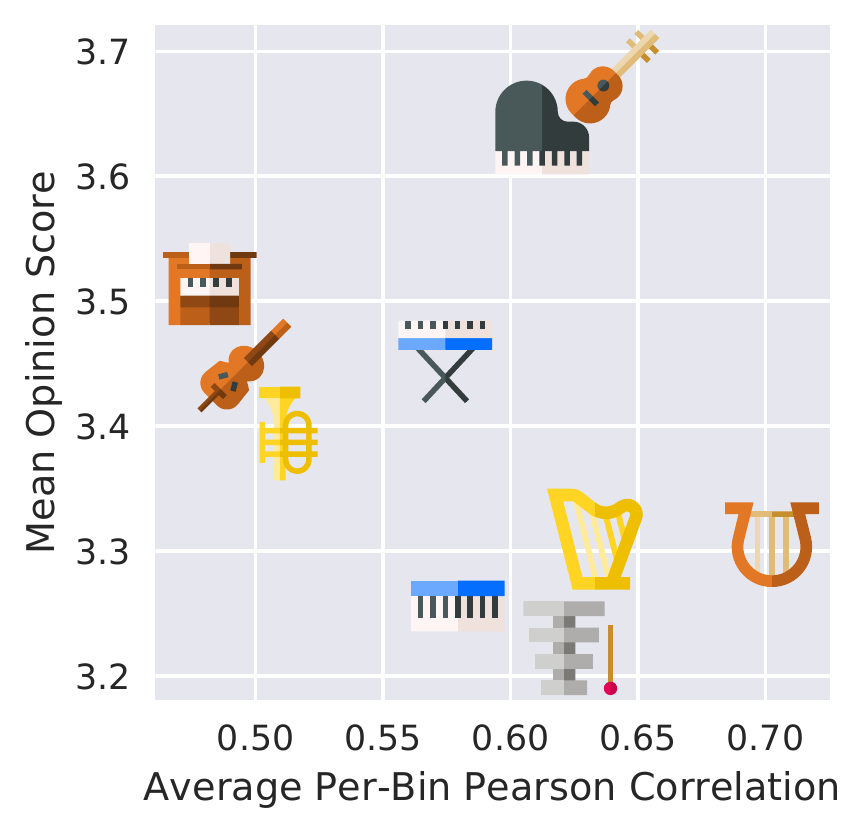}
	\caption{Per-instrument plot of synthesis quality, using the same icons as in Fig. \ref{fig:embedding}.\newline}
	\label{fig:scatter}
	\vspace{-1em}
\end{wrapfigure}

In Figure \ref{fig:scatter}, we compare the numerically and perceptually evaluated quality for each instrument.
The horizontal coordinates are the average values for each curve in Figure \ref{fig:correlations}b.
Pearson correlations between CQT are not necessarily indicative of the subjective synthesis quality, because a large contrast in the temporal envelope can contribute to high Pearson correlations, notwithstanding a low perceptual quality.
Reflecting this, more transient instruments in the lower right such as pizzicato strings achieve higher Pearson correlations compared to the MOS, while more sustained instruments on the left side have relatively lower Pearson correlations but have higher perceptual quality.

\subsection{The Timbre Embedding Space}

\begin{figure}[t]
	\centering
	\includegraphics[width=\columnwidth]{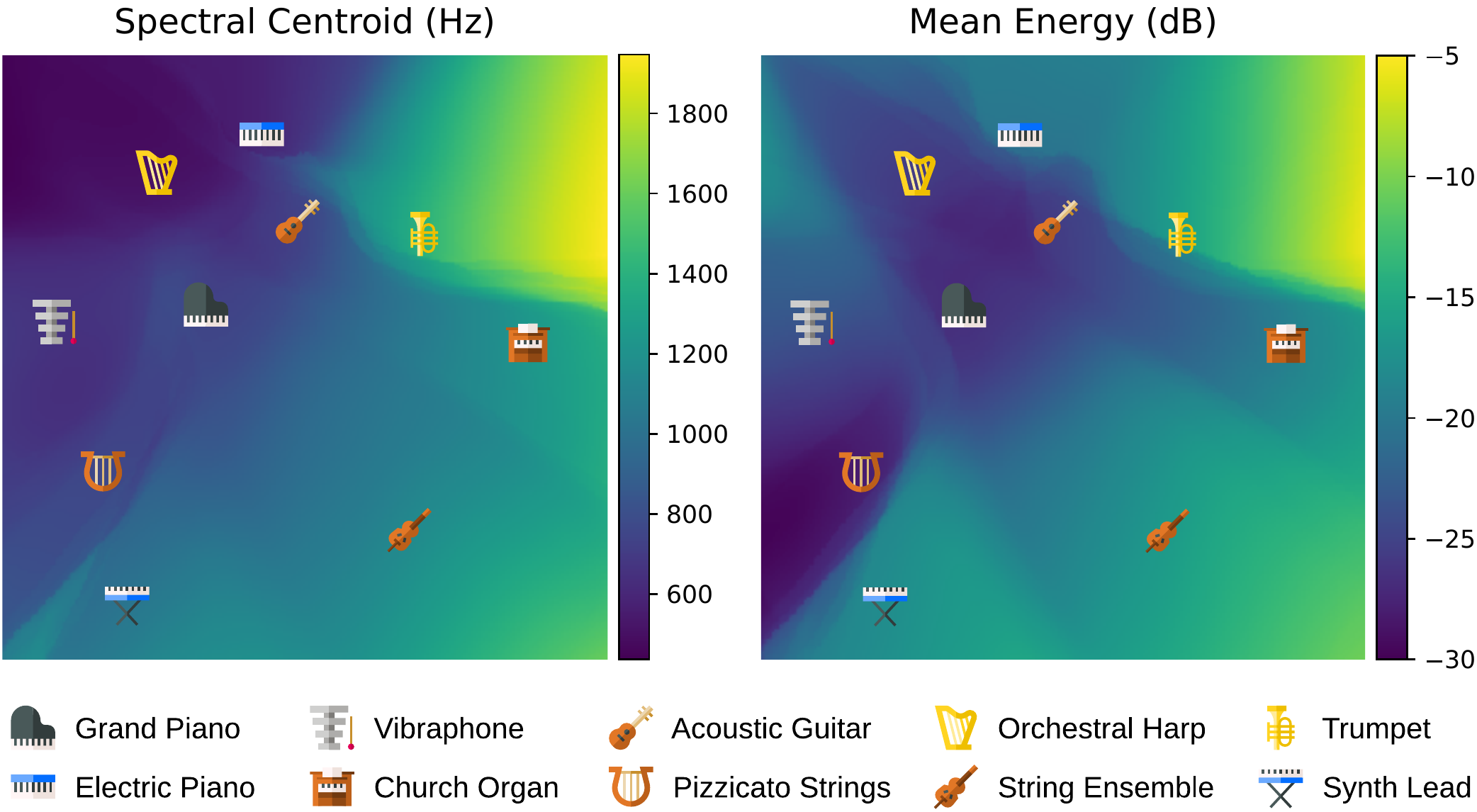}
	\caption{Visualization of the embedding space trained in 2 dimensions, using spectral centroids and mean energy. The continuous colors are obtained for each pixel in the 320-by-320 grid and are related to the spectral and temporal envelopes of the timbres.}
	\label{fig:embedding}
	\vspace{-1em}
\end{figure}

To make sense of how the learned embedding space conveys timbre information, we construct a 320-by-320 grid that encloses all instrument embeddings and predict the Mel spectrogram conditioned on every pixel in the grid.
The spectral centroid and the mean energy corresponding to each pixel are plotted in Figure \ref{fig:embedding}, which are indicative of the two main aspects of instrumental timbres: the spectral and temporal envelopes.
A higher spectral centroid signifies stronger harmonic partials in high frequency, while a lower spectral centroid indicates that it is closer to a pure sine tone.
Similarly, higher mean energy implies a more sustained tone, and low mean energy means that the note is transient and decays rather quickly.
The points corresponding to the 10 instruments are annotated with instrument icons.\footnote{The icons are made by Freepik and licensed by CC 3.0 BY.}
These plots show that the learned embedding space forms a continuous span over the timbres expressed by all instruments in the training data.
This allows us to use the timbre embedding as a flexible control space for the synthesizer, and timbre morphing is possible by interpolating along curves within the embedding space.

To illustrate how the model scales with more diverse timbre, we train the Mel2Mel model with 100 instruments using a 10-dimensional embedding, and we refer the readers to the web demo$^\textrm{5}$ for an interactive $t$-SNE visualization~\cite{maaten2008visualizing} of the embedding space.
The 10-dimensional embedding space also contains a locally continuous timbre distribution of instruments, as in Figure \ref{fig:embedding}, implying that the Mel2Mel model is capable of scaling to hundreds of instruments and to a higher-dimensional embedding space.

In addition to the audio samples used in the experiments, our interactive web demo\footnote{\url{https://neural-music-synthesis.github.io}} showcases the capability of flexible timbre control, where the Mel2Mel model runs on browser to convert preloaded or user-provided MIDI files into Mel spectrograms using a user-selected point in the embedding space.

\section{Conclusions and Future Directions}\label{sec:conclusions}

We showed that it is possible to build a music synthesis model by combining a recurrent neural network and FiLM conditioning layers, followed by a WaveNet vocoder.
It successfully learns to synthesize musical notes according to the given note sequence and timbre embedding in a continuous timbre space, providing the ability of flexible timbre control for music synthesizers.

The capacity of the WaveNet, such as the number of residual channels and the number of layers, is limited due to the memory requirements of the nv-wavenet implementation, and the degradation from $\mu$-law quantization is also apparent in the experiments.
These limitations can be overcome by Parallel WaveNet \cite{oord2017parallel}, which does not require a special CUDA kernel for fast synthesis and uses a continuous probability distribution for generation, thereby avoiding the quantization noise.
Our earlier experiments on continuous emission failed to stably perform autoregressive sampling due to teacher forcing, and the future work includes investigating this phenomenon comparing with \cite{anonymous2019maestro}, which used a mixture of logistics distributions to produce high-quality piano sounds.

A notable observation is that the WaveNet vocoder is able to synthesize polyphonic music from Mel spectrograms containing only 80 frequency bins, which are not even aligned to the tuning of the audio files.
While more information available from the increased bins should help synthesize more accurate audio, predicting the higher-dimensional representation becomes more compute-intensive and inaccurate, making 80 bins a sweet spot for use with WaveNet.
Introducing an adversarial loss function for predicting high-resolution images \cite{ledig2017srgan} can be a viable direction for predicting more accurate and realistic Mel spectrograms for conditioning WaveNet.

Overall, we have demonstrated that a MIDI-to-audio synthesizer can be learned directly from audio, and that this learning allows for flexible timbre control.
Once extended with an improved vocoder and trained on real audio data, we believe the model can result in a powerful and quite realistic music synthesis model.

\bibliographystyle{IEEEbib}
\bibliography{neural-music-synthesis}

\begin{thebibliography}{10}

\bibitem{pejrolo2017creating}
Andrea Pejrolo and Scott~B Metcalfe,
\newblock {\em {Creating Sounds from Scratch: A Practical Guide to Music
  Synthesis for Producers and Composers}},
\newblock Oxford University Press, 2017.

\bibitem{caetano2010morph}
Marcelo~Freitas Caetano and Xavier Rodet,
\newblock ``{Automatic Timbral Morphing of Musical Instrument Sounds by
  High-Level descriptors},''
\newblock in {\em Proceedings of the International Computer Music Conference},
  2010.

\bibitem{osaka1995timbre}
Naotoshi Osaka,
\newblock ``{Timbre Interpolation of Sounds Using a Sinusoidal Model},''
\newblock in {\em Proceedings of the International Computer Music Conference},
  1995.

\bibitem{boccardi2001sound}
Federico Boccardi and Carlo Drioli,
\newblock ``{Sound Morphing with Gaussian Mixture Models},''
\newblock in {\em Proceedings of DAFx}, 2001.

\bibitem{slaney1996automatic}
Malcolm Slaney, Michele Covell, and Bud Lassiter,
\newblock ``{Automatic Audio Morphing},''
\newblock in {\em Proceedings of the {IEEE} International Conference on
  Acoustics, Speech, and Signal Processing ({ICASSP})}, 1996, vol.~2.

\bibitem{ezzat2005morphing}
Tony Ezzat, Ethan Meyers, James Glass, and Tomaso Poggio,
\newblock ``{Morphing Spectral Envelopes using Audio Flow},''
\newblock in {\em European Conference on Speech Communication and Technology},
  2005.

\bibitem{caetano2013musical}
Marcelo Caetano and Xavier Rodet,
\newblock ``{Musical Instrument Sound Morphing Guided by Perceptually Motivated
  Features},''
\newblock {\em {IEEE} Transactions on Audio, Speech, and Language Processing},
  vol. 21, no. 8, 2013.

\bibitem{peeters2011timbre}
Geoffroy Peeters, Bruno~L Giordano, Patrick Susini, Nicolas Misdariis, and
  Stephen McAdams,
\newblock ``{The timbre Toolbox: Extracting Audio Descriptors from Musical
  Signals},''
\newblock {\em the Journal of the Acoustical Society of America}, vol. 130, no.
  5, 2011.

\bibitem{grey1977multidimensional}
John~M Grey,
\newblock ``{Multidimensional Perceptual Scaling of Musical Timbres},''
\newblock {\em the Journal of the Acoustical Society of America}, vol. 61, no.
  5, 1977.

\bibitem{wessel1979timbre}
David~L Wessel,
\newblock ``{Timbre Space as a Musical Control Structure},''
\newblock {\em Computer Music Journal}, 1979.

\bibitem{logan2000mfcc}
Beth Logan,
\newblock ``{Mel Frequency Cepstral Coefficients for Music Modeling},''
\newblock in {\em Proceedings of the International Society for Music
  Information Retrieval ({ISMIR}) Conference}, 2000.

\bibitem{agostini2003musical}
Giulio Agostini, Maurizio Longari, and Emanuele Pollastri,
\newblock ``{Musical instrument timbres classification with spectral
  features},''
\newblock {\em EURASIP Journal on Advances in Signal Processing}, vol. 2003,
  no. 1, 2003.

\bibitem{humphrey2011nlse}
Eric~J Humphrey, Aron~P Glennon, and Juan~Pablo Bello,
\newblock ``{Non-Linear Semantic Embedding for Organizing Large Instrument
  Sample Libraries},''
\newblock in {\em Proceedings of the International Conference on Machine
  Learning Applications ({ICMLA})}, 2011, vol.~2.

\bibitem{esling2018timbrevae}
Philippe Esling, Axel Chemla–Romeu-Santos, and Adrien Bitton,
\newblock ``{Generative Timbre Spaces with Variational Audio Synthesis},''
\newblock in {\em Proceedings of the International Conference on Digital Audio
  Effects ({DAFx})}, 2018.

\bibitem{kingma2014vae}
Diederik~P. Kingma and Max Welling,
\newblock ``{Auto-Encoding Variational Bayes},''
\newblock in {\em Proceedings of the International Conference on Learning
  Representations ({ICLR})}, 2014.

\bibitem{oord2016wavenet}
Aaron van~den Oord, Sander Dieleman, Heiga Zen, Karen Simonyan, Oriol Vinyals,
  Alex Graves, Nal Kalchbrenner, Andrew Senior, and Koray Kavukcuoglu,
\newblock ``{WaveNet: A Generative Model for Raw Audio},''
\newblock {\em arXiv:1609.03499}, 2016.

\bibitem{shen2018tacotron}
Jonathan Shen, Ruoming Pang, Ron~J Weiss, Mike Schuster, Navdeep Jaitly,
  Zongheng Yang, Zhifeng Chen, Yu~Zhang, Yuxuan Wang, Rj~Skerrv-Ryan, et~al.,
\newblock ``{Natural TTS Synthesis by Conditioning WaveNet on Mel Spectrogram
  Predictions},''
\newblock in {\em Proceedings of the {IEEE} International Conference on
  Acoustics, Speech, and Signal Processing ({ICASSP})}, 2018.

\bibitem{ping2018deepvoice}
Wei Ping, Kainan Peng, Andrew Gibiansky, Sercan~O. Arik, Ajay Kannan, Sharan
  Narang, Jonathan Raiman, and John Miller,
\newblock ``{Deep Voice 3: 2000-Speaker Neural Text-to-Speech},''
\newblock in {\em Proceedings of the International Conference on Learning
  Representations ({ICLR})}, 2018.

\bibitem{engel2017neural}
Jesse Engel, Cinjon Resnick, Adam Roberts, Sander Dieleman, Mohammad Norouzi,
  Douglas Eck, and Karen Simonyan,
\newblock ``{Neural Audio Synthesis of Musical Notes with WaveNet
  Autoencoders},''
\newblock in {\em Proceedings of the International Conference on Machine
  Learning ({ICML})}, 2017, vol.~70.

\bibitem{mor2018universal}
Noam Mor, Lior Wolf, Adam Polyak, and Yaniv Taigman,
\newblock ``{A Universal Music Translation Network},''
\newblock {\em arXiv:1805.07848}, 2018.

\bibitem{anonymous2019maestro}
Anonymous,
\newblock ``{Enabling Factorized Piano Music Modeling and Generation with the
  MAESTRO Dataset},''
\newblock {\em Submitted to the International Conference on Learning
  Representations ({ICLR})}, 2019,
\newblock under review.

\bibitem{hawthorne2017onsets}
Curtis Hawthorne, Erich Elsen, Jialin Song, Adam Roberts, Ian Simon, Colin
  Raffel, Jesse Engel, Sageev Oore, and Douglas Eck,
\newblock ``{Onsets and frames: Dual-Objective Piano Transcription},''
\newblock in {\em Proceedings of the International Society for Music
  Information Retrieval ({ISMIR}) Conference}, 2018.

\bibitem{perez2018film}
Ethan Perez, Florian Strub, Harm De~Vries, Vincent Dumoulin, and Aaron
  Courville,
\newblock ``{Film: Visual Reasoning with a General Conditioning Layer},''
\newblock in {\em Proceedings of the Association for the Advancement of
  Artificial Intelligence ({AAAI}) Conference}, 2018.

\bibitem{thickstun2017musicnet}
John Thickstun, Zaid Harchaoui, and Sham Kakade,
\newblock ``{Learning Features of Music from Scratch},''
\newblock in {\em Proceedings of the International Conference on Learning
  Representations ({ICLR})}, 2017.

\bibitem{emiya2010multipitch}
Valentin Emiya, Roland Badeau, and Bertrand David,
\newblock ``{Multipitch Estimation of Piano Sounds using a New Probabilistic
  Spectral Smoothness Principle},''
\newblock {\em {IEEE} Transactions on Audio, Speech, and Language Processing},
  vol. 18, no. 6, 2010.

\bibitem{itu2011mushra}
``{Method for the Subjective Assessment of Intermediate Quality Level of Coding
  Systems},''
\newblock in {\em ITU-R Recommendation BS.1534-1}, 2001.

\bibitem{oord2017parallel}
Aaron van~den Oord, Yazhe Li, Igor Babuschkin, Karen Simonyan, Oriol Vinyals,
  Koray Kavukcuoglu, George van~den Driessche, Edward Lockhart, Luis~C Cobo,
  Florian Stimberg, et~al.,
\newblock ``{Parallel WaveNet: Fast High-Fidelity Speech Synthesis},''
\newblock {\em arXiv:1711.10433}, 2017.

\bibitem{maaten2008visualizing}
Laurens van~der Maaten and Geoffrey Hinton,
\newblock ``{Visualizing Data Using t-SNE},''
\newblock {\em Journal of Machine Learning Research}, vol. 9, no. Nov, 2008.

\bibitem{ledig2017srgan}
Christian Ledig, Lucas Theis, Ferenc Husz{\'a}r, Jose Caballero, Andrew
  Cunningham, Alejandro Acosta, Andrew~P Aitken, Alykhan Tejani, Johannes Totz,
  Zehan Wang, et~al.,
\newblock ``{Photo-Realistic Single Image Super-Resolution Using a Generative
  Adversarial Network.},''
\newblock in {\em Proceedings of the Conference on Computer Vision and Pattern
  Recognition ({CVPR})}, 2017, vol.~2.

\end{thebibliography}

\end{document}